\documentclass[reprint, amsmath,amssymb,twocolumn, prl]{revtex4} 
\usepackage{url}
\usepackage[english]{babel}

\usepackage{chemist}
\usepackage{blindtext}
\usepackage{lipsum}
\usepackage{graphicx}

\begin{document}

\title{Atomistic Mechanisms for the Nucleation of Aluminium Oxide Nanoparticles}

\author{Julien Lam}
\email{julien.lam@univ-lyon1.fr}
\author{David Amans}
\email{david.amans@univ-lyon1.fr}
\author{Christophe Dujardin}
\author{Gilles Ledoux}
\author{Abdul-Rahman Allouche}

\affiliation{Universit\'e Lyon 1, F-69622 Villeurbanne, France, UMR5306 CNRS, Institut Lumiere Matiere, PRES-Universit\'e de Lyon, F-69361 Lyon, France}

\begin{abstract}
A predictive model for nanoparticle nucleation has not yet been successfully achieved. Classical nucleation theory fails because the atomistic nature of the seed has to be considered since geometrical structure as well as stoichiometry do not always match the bulk values. We present a fully microscopic approach based on a first-principle study of aluminium oxide clusters. We have calculated stable structures of \chemform{Al_xO_y} and their associated thermodynamic properties. From these data, the chemical composition of a gas composed of aluminium and oxygen atoms can be calculated as a function of temperature, pressure, and aluminium to oxygen ratio.
We demonstrate the accuracy of this approach in reproducing experimental results obtained with time resolved  spectroscopy of a laser induced plasma from an \chemform{Al_2O_3} target. We thus extended the calculation to lower temperatures, \textit{i.e.} longer time scales, to propose a scenario of  composition gas evolution  leading to the first alumina seeds.
\end{abstract}

\maketitle
\section{Introduction}

Although the nucleation of nanoparticles is of crucial interest for a wide range of applications, such as the nanoparticle synthesis or the fight against atmospheric nanoparticles, a predictive model of its mechanisms is not yet achieved.
Nucleation consists on the formation of a more stable seed within a metastable phase. This phenomenon can be observed in any first-order phase transition especially with vapour-phase condensation, liquid-to-vapour boiling, solid-state precipitation and binary separations~\cite{Ford2004, Clouet2010,Abraham1973,Schmelzer2005}. Since the beginning of the 19th century, classical nucleation theory (CNT) has been used for intuitive descriptions~\cite{Volmer1926,Farkas1927,Becker1935}. The CNT describes the formation  of a spherical particle immersed in a mother phase as a competition between the volume and the surface energies. While $\Delta G_v$ the standard Gibbs free energy difference per unit volume  between the two phases favors the growth, the surface term, proportional
to the surface tension $\sigma$, has an opposite effect. As a result, CNT states the existence of a minimum radius $r^{\star}$ allowing the new phase to grow for radii larger than $r^{\star}$. $r^{\star}$ is defined by~\cite{Abraham1973}: 
\begin{equation}
r^{\star}=-\frac{2\sigma}{\Delta G_v},
\end{equation}
CNT appears to be successful for liquid nucleation in a supersaturated vapour~\cite{Strey1994, Diemand2014}. However, this approach remains controversial regarding quantitative results~\cite{Viisanen1993,Viisanen1994,Hruby1996,Sen1999}, and its conceptual limitations have been raised by several authors~\cite{Viisanen1993,Sen1999,Lutsko2013,Lutsko2015}.
On the one hand, CNT assumes that the seed is homogeneous and isostructural to the bulk crystal.  
On the other hand, the Otswald rule~\cite{Ostwald1897,Threlfall2003,Zhang2009} states that the first growing phase is not always the most stable. During the crystal growth, the geometric structure may indeed reorganise with transient states potentially having a different structure from the bulk analog~\cite{Baletto2000,Baletto2001}. 
A second limitation of CNT arises from the capillary approximation which assumes that nucleus and bulk materials have the same thermodynamic properties, and particularly the same surface tension. As an illustration, one may consider the case of \chemform{Al_2O_3}. For both crystallographic phases ($\alpha$ or $\gamma$), $\sigma$ is almost 1.5~J.m$^{-2}$ for nanoparticles~\cite{McHale1997} and $\Delta G_v$ is almost -62~kJ.cm$^{-3}$ at 298~K~\cite{NIST}. CNT predicts a critical radius of about the Bohr radius. 
Lastly, the nucleation rate is driven by the probability of a seed reaching $r^{\star}$ by addition of molecules~\cite{Volmer1926,Farkas1927}.
This approach involves parameters difficult to quantify~\cite{Shore2000}. Consequently, the CNT usually fails to predict the nucleation rate, with tens of orders of magnitude discrepancy between theory and experiment~\cite{Sen1999}.  

Over and beyond the CNT's limitations, the need to explore crystal nucleation becomes increasingly crucial since it may provide a control on the crystal structure and size distribution for nanoparticle synthesis \cite{Erdemir2009}. Experimentally, it is challenging to probe nucleation processes since it involves time and length scales that are usually too fast and too small in most experiments. 

For computer simulations, two trends can be identified in the literature. On the one hand, molecular dynamics calculations are used with hard-sphere~\cite{Williams2008}, Lennard-Jones~\cite{Diemand2014,Lummen2004, Chakraborty2013} or more refined potentials~\cite{Shen2013,Desgranges2009,Shibuta2011,Shore2000} to investigate the nucleation kinetics. On the other hand, first principle calculations are used to study with precision clusters for particular systems such as zinc \cite{Sunaidi2008}, silicium \cite{Bromley2007}, and titanium \cite{Calatayud2008,Qu2006} oxides. Apart from the work of Loschen~\textit{et~al.}~\cite{Loschen2007} who tested different stoichiometries of cerium oxides only the bulk stoichiometry is generally considered in simulations. In addition, very few works compared these computational results with experimental measurements~\cite{Burnin2002}.

Our work is carried out in the context of laser ablation, including pulsed laser deposition~\cite{phipps2007} or pulsed laser ablation in liquids~\cite{Barcikowski2013,Lam2014}. But it can be generalized to other methods such as cluster sources~\cite{DeHeer1993} and gas-phase combustion~\cite{gardiner2000}. For all these techniques, an atomic vapour or a plasma,  is generated at temperatures higher than a thousand Kelvins. Then, the system is quenched down to room temperature, leading to nucleation and growth of the nanoparticles. We investigate nucleation processes through an original use of different tools from quantum chemistry combined with experimental measurements obtained from laser induced plasma spectroscopy~\cite{Omenetto2010}.  We choose to work with a model system made of aluminium and oxygen atoms. We first compute the lowest energy structures of \chemform{Al_xO_y} molecules. Thereafter, we calculate the gas-phase equilibrium composition as a function of the temperature, the pressure and the initial ratio of aluminium to oxygen atoms. For the highest temperatures, the computed composition is compared to experimental measurements obtained from the laser ablation of an alumina target ($\alpha$-\chemform{Al_2O_3}). Moreover, plasma spectroscopy allows the probing of short time scales, when the plasma is hot and optically active.

\section{Numerical calculations}

\begin{figure}[!ht]
\begin{center}
\includegraphics[width=8.5cm]{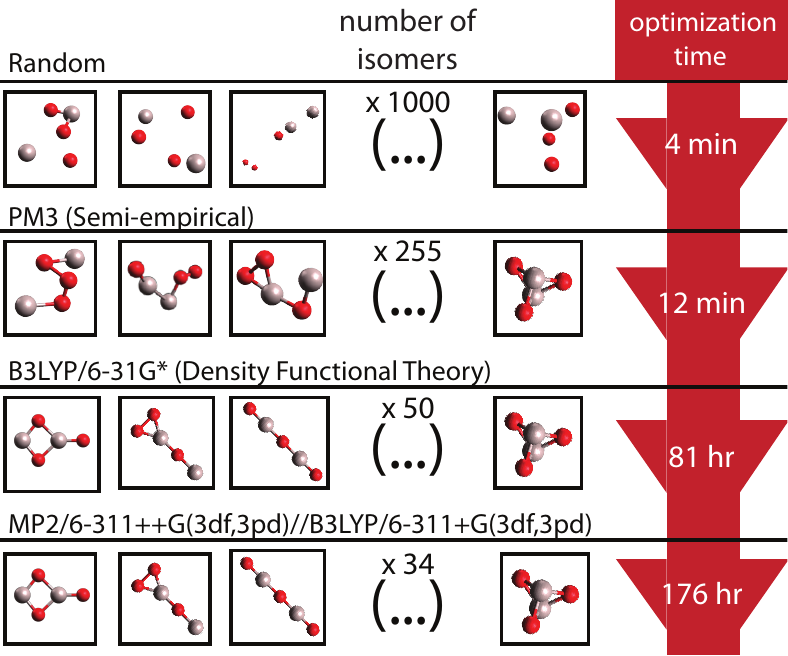}
\caption{Computational optimization algorithm illustrated by the example of \chemform{Al_2O_3} molecules. After each step, similar geometries are removed. The number of remaining geometries is indicated above the brackets. The time corresponds to the computational time for each step.}
\label{Fig:CompScheme}
\end{center}
\end{figure}

Figure~\ref{Fig:CompScheme} reviews schematically the computational algorithm used for the structural investigation. For each value of x and y, we start from a set of 1000 geometries where atoms are randomly disposed with interatomic distances corresponding to the covalent bond distances. The system is then relaxed via a PM3 semi-empirical method \cite{Stewart1989}. The remaining geometries are then optimized using Density Functional Theory (DFT) calculation. Two sets of bases are used successively, B3LYP/6-31G* and B3LYP/6-311+G(3df,3pd).
The Gibbs energies were determined using MP2 and B3LYP according to the following steps. The harmonic frequencies were calculated by B3LYP/6-311+G(3df,3pd) using the structures optimized. MP2 thermochemistry was determined by adding B3LYP thermal correction factors to the MP2 single-point energies and is reported as MP2/6-311++G(3df,3pd)//B3LYP/6-311+G(3df,3pd).

The dissociated geometries are removed and only geometries whose energy is at most 2~eV higher than the ground state are kept as the others are not relevant for the temperatures studied . 
For the biggest clusters \chemform{(Al_2O_3)_3} and \chemform{(Al_2O_3)_4}, the structures published by Sharipov~\textit{et~al.}~\cite{Sharipov2013} were used as inputs in our optimization process.
DFT and MP2 calculations were performed with the Gaussian09 D01 revision.\cite{g09}

We investigated all the molecular formulas following $(x,y) \in [1;4]$. In addition, two stoichiometric trends were followed, \chemform{(AlO)_n} and \chemform{(Al_2O_3)_n}. \chemform{(AlO)_n} with $n \leqslant 8$ were chosen because Patzer~\textit{et~al.} demonstrated that for clusters with x and y smaller than 4, this stoichiometry is the most stable~\cite{Patzer2005}.  \chemform{(Al_2O_3)_n} with $n \leqslant 4$ were investigated because it corresponds to the bulk stoichiometry. 

\begin{figure}[!ht]
\begin{center}
\includegraphics[width=8.5cm]{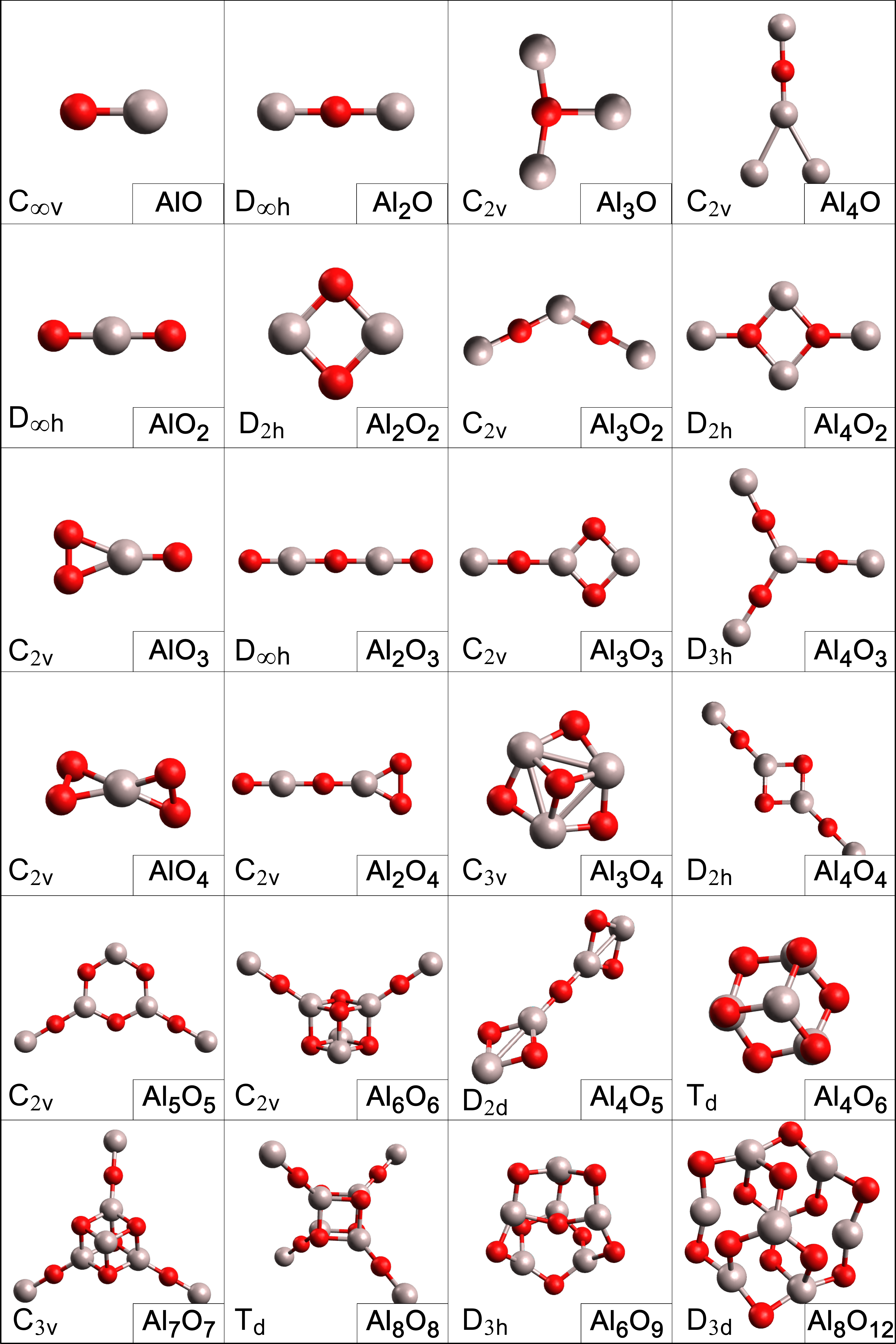}
\caption{Most stable structures obtained at MP2/6-311++G(3df,3pd)//B3LYP/6-311+G(3df,3pd) level of theory. The grey spheres and the red spheres correspond respectively to the aluminum atoms and the oxygen atoms.}
\label{Fig:MoleculeMostStable}
\end{center}
\end{figure}
The figure~\ref{Fig:MoleculeMostStable} shows the most stable structures we obtained. Point-zero energies and geometrical parameters can be found in the Supporting Information (see SI.1). The method we used provides a systematic procedure to select the different stable chemical structures for a given molecular formula. Indeed, for some of the molecules, the results are consistent with various works published previously \cite{Patzer2005, Li2012, Sharipov2013}. Nevertheless, for others such as \chemform{AlO_4} and \chemform{Al_2O_3}, we were able to find isomers that are more stable than those obtained earlier.

The first principle calculations provide dissociation energy, vibrational and rotational constants for all \chemform{Al_x O_y^{(i)}} molecules, where $i$ denotes an isomer of the molecular formula \chemform{Al_x O_y}. 
The formation Gibbs free energy $\Delta_f G_{Al_x O_y^{(i)}}$ of each \chemform{Al_xO_y^{(i)}} is then computed as a function of the temperature $T$ and the pressure $P_\circ$. The Gibbs free energy calculation is described in the Supporting Information SI.2.
Based on the model proposed by Patzer \textit{et al.}~\cite{Patzer2005}, we computed the composition of a gas fed with aluminium and oxygen atoms as a function of the temperature, the initial proportion of elements ($\lambda\equiv N_{Al}/N_{O}$) and the pressure.
We considered the set of reactions which correspond to the formation of the \chemform{Al_xO_y^{(i)}} molecules from an atomic gas: 
\begin{chemeqn}
\label{ElementReaction}
x Al + y O \rightleftarrows Al_x O_y^{(i)}
\end{chemeqn}
The gibbs free energy of reactions $\Delta_r G_{Al_x O_y^{(i)}}$ is deduced from the $\Delta_f G_{Al_x O_y^{(i)}}$~\cite{McQuarrie1999}. 
We improved the Patzer's model by taking into account the temperature and pressure dependences of $\Delta_r G_{Al_x O_y^{(i)}}$  and the contribution of all isomers for each molecules (see Supporting Information SI.2).

\begin{figure}[!ht]
\begin{center}
\includegraphics[width=8.5cm]{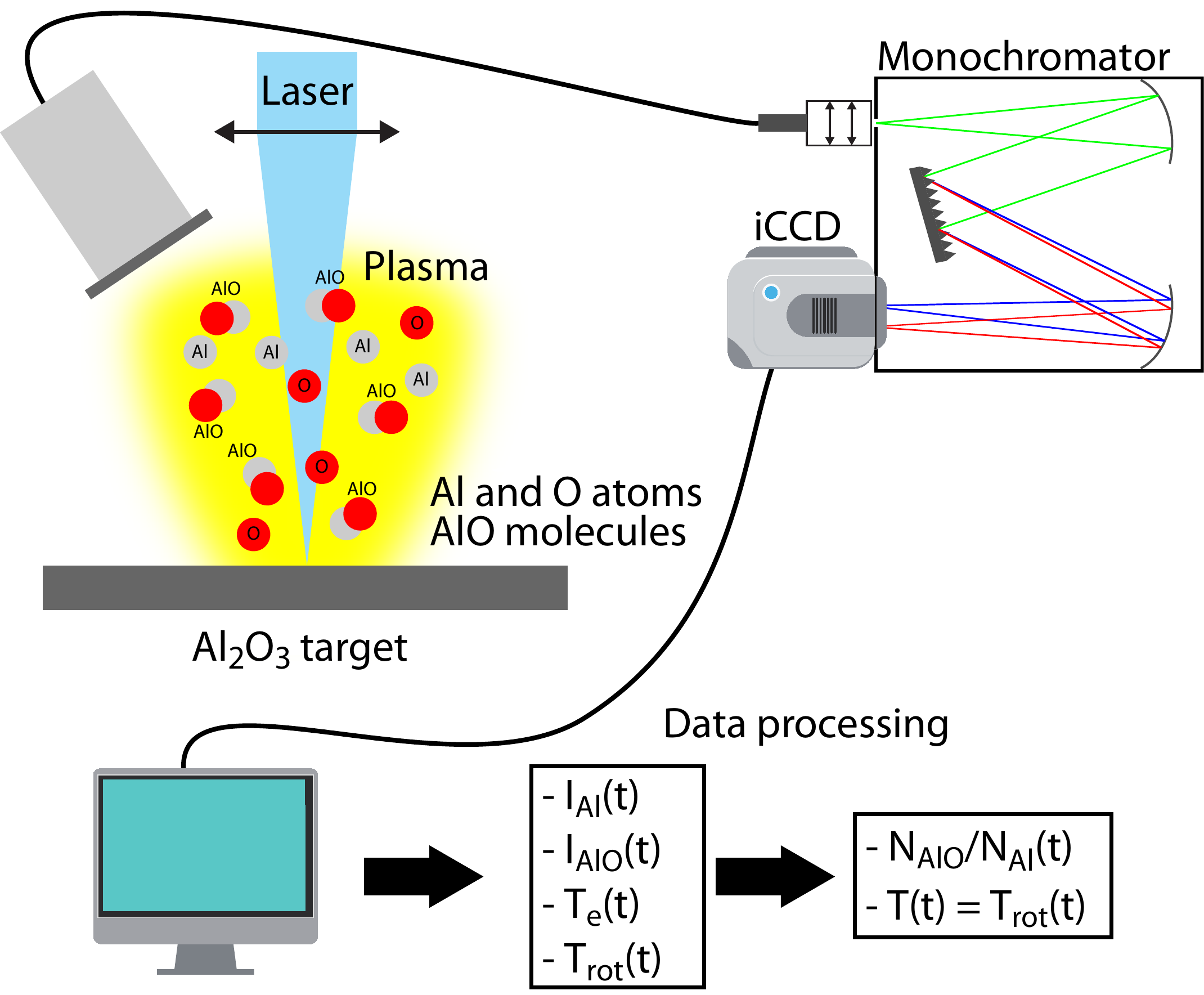}
\caption{Sketch summarizing the experimental method: (i) a pulsed-laser is focused onto an \chemform{Al_2O_3} solid target generating an optically active plasma, (ii) the emitted light is recorded using a monochromator and an intensified CCD, (iii) Al and AlO intensities along with $T_e$ and $T_{rot}$ as a function of the time, (iv) intensities are corrected leading to a ratio in density.}
\label{Fig:ExpScheme}
\end{center}
\end{figure}

\section{Experimental data}

The experimental measurements used for this work have been published previously~\cite{Lam2014b} and are sketched in figure~\ref{Fig:ExpScheme}. To summarize, we performed the spectroscopic characterisation of a plasma induced by the laser ablation of an \chemform{Al_2O_3} target in ambient air. The emission intensity of aluminium monoxide molecules and aluminium atoms was measured as a function of time. The electronic temperature $T_e$ was also measured~\cite{Lam2014b}. Using the formalisms developed in our previous article~\cite{Lam2014}, we deduced the density ratio $N_{AlO}/N_{Al}$. In the meantime, we measured a temporal evolution of the rotational temperature $T_{rot}$. This can account for the temperature used in the thermochemistry model because the rotational temperature probes the kinetic temperature of atoms~\cite{Lam2014b}.  $N_{AlO}/N_{Al}$ is then reported in figure~\ref{Fig:Map-ThvsExp}(a) as a function of the time (bottom axis), but also as a function of the temperature $T_{rot}$ (top axis). These experimental data will be used to validate the thermochemistry model.

The  $\lambda$ ratio is expected to be between two extreme values. $\lambda_{max}=2/3$ corresponds to a plasma only composed by the ablated matter. $\lambda_{min}$ corresponds to the ablated matter combined with the ambient air in the same plasma plume volume.
The amount of matter ablated is obtained from the crater shape.
The crater depth is measured using Alpha-Step D100 profiler from Tencor. 
The crater depth measured after 5 pulses is 1.5~$\mu$m~$\pm$~500~nm.
The crater diameter is measured with an optical microscope. 
The crater diameter is 500~$\mu$m~$\pm$~100~$\mu$m.
Assuming an $\alpha$-Al$_2$0$_3$ density of 3.95~g.cm$^{-3}$, a molar mass of 101.96 g.mol$^{-1}$, we have obtained $7\times 10^{15}$~$\pm$~70\% atoms ablated per pulse. 
Considering the size of the plasma, 2~mm in diameter~\cite{Lam2014b}, and assuming an ideal gas, the number of \chemform{O_2} molecules contained in the same volume of air is $2\times 10^{16}$. 
The ratio between the number of aluminium atoms provided by the target and the number of oxygen atoms provided by the target and the air leads to a $\lambda_{min}$ of 0.06 (see Supporting Information SI.3).

\section{Discussion}

\begin{figure}[!ht]
\begin{center}
\includegraphics[width=8.5cm]{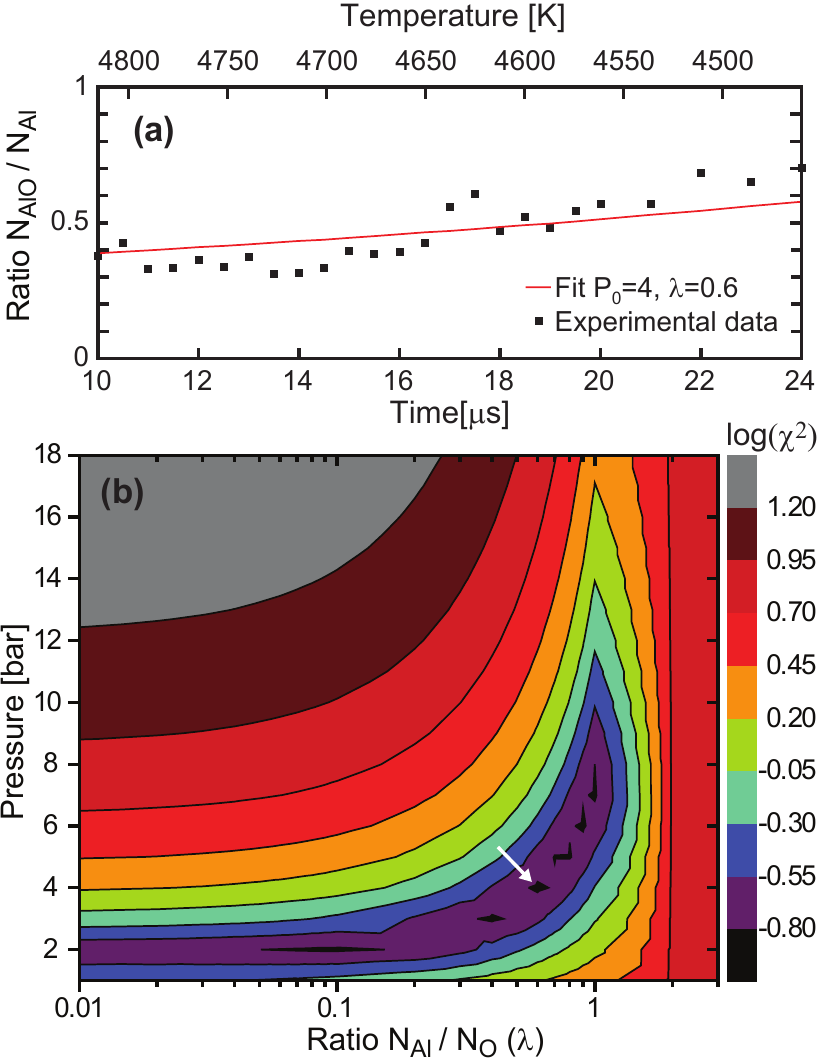}
\caption{
(a) The black squares correspond to the ratio $N_{AlO}/N_{Al}$ measured as a function of time (bottom axis) and temperature (top axis).
The solid line corresponds to the best fit, for $P_{\circ}=4$~bars and $\lambda=N_{Al}/N_{O}$=0.6, deduced from the smallest figure of merit reported in panel (b).
Numerical results for different values  of $\lambda$ and $P_{\circ}$ are shown in the Supporting Information SI.4.
(b) Likelihood between the experimental ratio $N_{AlO}/N_{Al}$ reported in the panel (a) and the theoretical ratio, as a function of the ambient pressure $P_{\circ}$ and the stoichiometry $\lambda$. The colorbar shows the $\chi ^2$ values in logarithmic scale. The white arrow shows the minimum value of the $\chi ^2$. 
}
\label{Fig:Map-ThvsExp}
\end{center}
\end{figure}

In figure~\ref{Fig:Map-ThvsExp}(b), the experimental measurements from 10~$\mu$s to 24~$\mu$s are compared with our calculations using the Pearson's cumulative statistic test($  \chi ^2 = \sum \left( X_{th} - X_{exp}\right) ^2 $). We emphasis the valley of highest likelihood which exhibits a minimum of $\chi ^2$ for $P_{\circ}=4$~bars and $\lambda=0.6$. The $\lambda$ value is consistent with the expected one. The deduced pressure is also consistent with previous plasma spectroscopy measurements~\cite{Ma2010}.
Assuming these values, the theoretical ratio $N_{AlO}/N_{Al}$ is computed as a function of the temperature and compared to the experimental data in figure~\ref{Fig:Map-ThvsExp}(a). 
The agreement between theoretical calculation and experiment data relies on the idea that the plasma follows a quasi-static cooling in the considered time-range. The cooling rate measured is 25~K.$\mu$s$^{-1}$. The known kinetics of the chemical reactions leading to oxide molecules~\cite{LePicard1997} are fast enough to follow this cooling rate. 
This result is also consistent with the absence of an energy barrier for the reactions Al+O$\rightarrow$AlO and Al+O$_2 \rightarrow$AlO+O~\cite{Pak2003}. However, in order to assert equilibrium in the reacting gas, it could be relevant to take into account the homogeneity and the diffusion properties of the gas.

\begin{figure}[!ht]
\begin{center}
\includegraphics[width=8.5cm]{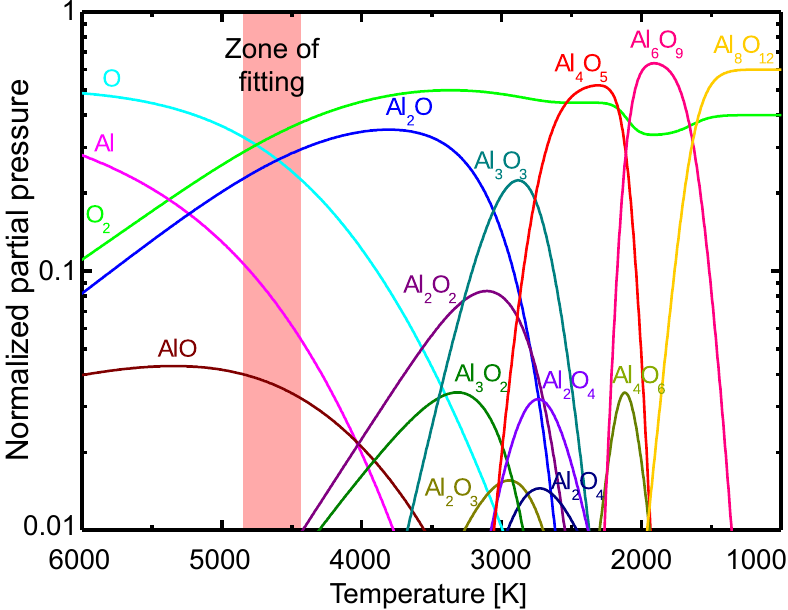}
\caption{Evolution of the gas-phase composition from 6000~K to 1000~K considering $P_{\circ}=4$~bars and $\lambda$=0.6. For clarity, the total pressure is normalized to one for each temperature. The partial pressure $P_{Al_xO_y}$ for a given $x$ and a given $y$ corresponds to the summation over all isomers.}
\label{Fig:CompoVsT}
\end{center}
\end{figure}

Based on the consistency between experimental and numerical results, we extended the thermochemistry calculation to lower temperature considering $P_{\circ}=4$~bars and $\lambda$=0.6. Figure~\ref{Fig:CompoVsT} shows the gas composition as a function of the temperature. Here, the partial pressure $P_{Al_xO_y}$ for a given molecular formula corresponds to the summation over all isomers partial pressure.
As expected, at high temperature ($T\gtrsim 5000\,K$) \textit{i.e.} short times, the gas is mainly composed of the smallest species \textit{i.e.} \chemform{Al}, \chemform{O}, \chemform{O_2}, \chemform{AlO} and \chemform{Al_2O}. Later, the amount of \chemform{Al_2O} and O simultaneously decrease leading to the two derivatives \chemform{Al_2O_2} and \chemform{Al_3O_3}. We emphasize that \chemform{Al_2O_3} never seems to emerge at this stage. 
The stoichiometry of alumina, corresponding to \chemform{Al_6O_9} and \chemform{Al_8O_{12}} molecules, emerges only for temperatures lower than 2000~K.
We observe a drastic dependence of the final composition for $\lambda$ varying around 2/3. It indicates that a small excess of oxygen is required to favor the (2:3) stoichiometry. This result is consistent with the general idea that to synthesize oxides, it is required to have a gas sursaturated in oxygen. 
For example, plasma enhanced chemical vapor deposition (PECVD) uses an oxygen plasma to
grow layers of oxides~\cite{Callard1999,Martinet1997}.
Pulsed laser deposition (PLD) of Al$_2$O$_3$ has been performed in an oxygen gas atmosphere to enhance the stoichiometry of the layer~\cite{Pillonnet2011}.
O$_2$ is injected during the growth of ZnO nanoparticles in low energy cluster beam deposition experiment (LECBD)~\cite{Tainoff2008}.

\begin{figure}[!ht]
\begin{center}
\includegraphics[width=8.5cm]{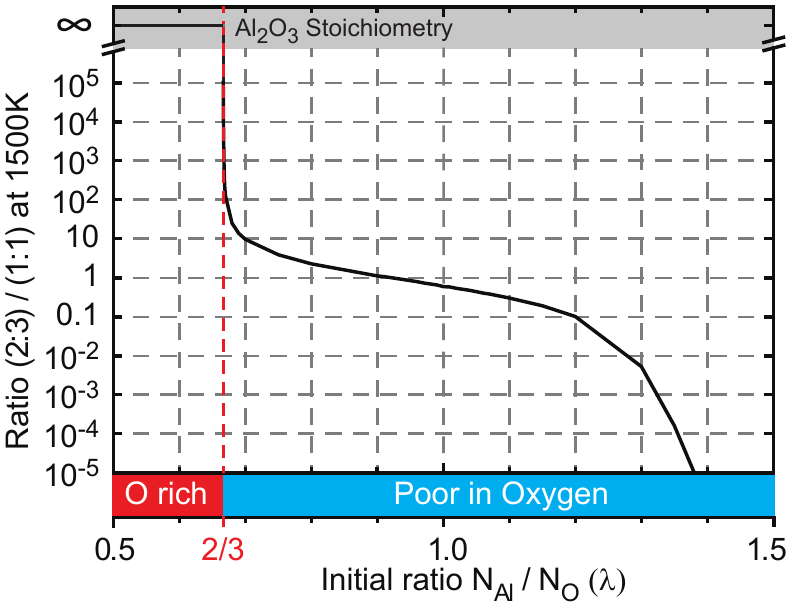}
\caption{Ratio in pressure between molecules following \chemform{(Al_2O_3)_n} and those following \chemform{(AlO)_n} as a function of $\lambda$ calculated at 1500~K. The full curve is shown in Supporting Information SI.5.}
\label{Fig:VsLambda}
\end{center}
\end{figure}

\section{Conclusions}

In summary, first principle calculations were employed to investigate aluminium oxide clusters at different stoichiometries. A systematic optimization approach was followed to obtain the stable structures. We find more stable isomers than the previously published ones for \chemform{AlO_4} and \chemform{Al_2O_3}  molecules~\cite{Patzer2005}. Temperature evolution of the composition of a gas made of aluminium and oxygen atoms was also calculated using these molecular properties. 
Although for high temperature, \chemform{(AlO)_n} is the most preponderant structure, the bulk aluminium oxide stoichiometry (2:3) starts to exceed the (1:1) stoichiometry for temperature lower than 2000~K. Besides the equilibrium considerations, the question of the kinetic of chemical reactions is not addressed here. Indeed, the occurrence of an equilibrium condition has to be combined with reasonable time scales of reaction kinetics to ensure the molecule formation. Especially, at low temperature, plasma spectroscopy can no longer provide the chemical composition since the system does not emit light. Nevertheless, the theoretical absorption and emission spectra can be deduced from our first-principles calculations. Laser induced fluorescence will be performed to probe the gas. Finally, the high level of theory employed in our calculations prevents addressing bigger clusters. A complementary work could consist on using the clusters we obtained to parametrize a semi-empirical model and perform molecular dynamics simulations. We were nevertheless able to predict the requirements for an oxygen rich gas for synthetizing the desired oxide stoichiometry.

\section{Supporting Information}
The geometry of each isomer is reported in the Supporting Information file \textit{Supplementary-xyz.zip} (See SI.1 for content description).
SI.2 gives the formulas used to calculate the Gibbs free energy and the gas composition.
SI.3 describes $\lambda_{min}$ calculation. 
The ratio $N_{AlO}/N_{Al}$ is computed for different values of $\lambda$ and $P_{\circ}$ in SI.4.
The full curve corresponding to the Figure~\ref{Fig:VsLambda} is shown in SI.5.

\section{Acknowledgement}
This work was granted access to the HPC resources of the FLMSN, "F\'ed\'eration Lyonnaise de Mod\'elisation et Sciences Num\'eriques", partner of EQUIPEX EQUIP@MESO.
The authors are grateful to Nora Abdellaoui for her participation on target crater measurements.

\bibliography{biblio}

\end{document}